\documentclass[showpacs,twocolumn,prl]{revtex4}

\usepackage{amssymb}
\usepackage{amsmath}
\usepackage{epsfig}

\begin{document}

\title{Generalized Bose-Einstein phase transition in large-$m$ component
 spin glasses}
\author{T. Aspelmeier}
\affiliation{Department of Physics and Astronomy, University of
Manchester, Manchester M13 9PL, UK}
\author{M. A. Moore}
\affiliation{Department of Physics and Astronomy, University of
Manchester, Manchester M13 9PL, UK}
\date{\today}

\begin{abstract}
  It is proposed to understand finite dimensional spin glasses using a $1/m$
  expansion, where $m$ is the number of spin components. It is shown that this
  approach predicts a replica symmetric state in finite dimensions.  The point
  about which the expansion is made, the infinite-$m$ limit, has been studied
  in the mean-field limit in detail and has a very unusual phase transition,
  rather similar to a Bose-Einstein phase transition but with $N^{2/5}$
  macroscopically occupied low-lying states.
\end{abstract}

\maketitle

After almost three decades of research, the nature of the low temperature
phase of finite dimensional spin glasses is not understood. The usual approach
is to start from the exactly soluble Sherrington-Kirkpatrick (SK) mean field
model and expand about it towards finite dimensions. This method resulted in
the monumental replica field theory, summarized in \cite{DeDominicisEtAl98} by
three of its main contributors. The results of this theory are that replica
symmetry is broken in finite dimensions, just as in the SK limit, down to 6
dimensions, below which all known calculational tools break down and not much
is known analytically. Yet there exists a mathematical proof that this picture
cannot hold in any finite dimension \cite{NewmanStein03}. This proof is
unfortunately non-constructive, so it does not give any insight into the
nature of the spin glass phase and it cannot decide between alternative
scenarios such as the droplet or the chaotic pairs picture. Neither does it
show where the replica field theory goes wrong. We therefore propose an
alternative method to investigate the finite dimensional spin glass phase in
order to bypass conventional replica field theory and to obtain a theory which
does not contradict the exact mathematical results.  To this end we shall
expand about the infinite component limit of the $m$-component spin glass in a
power series in $1/m$. As we will see, an expansion of this type preserves the
replica symmetry found in the $m=\infty$ model and will therefore give a
picture of the spin glass phase which is quite different to that of the usual
replica field theory. As part of our programme it is necessary to analyse the
$m=\infty$-component SK spin glass phase. The low temperature phase of this
model is a generalisation of a Bose-Einstein condensation in the sense that
the spins condense into a $n_0$-dimensional subspace of the $m$-dimensional
space they can occupy, where $n_0$ is a number of order $N^{2/5}$
\cite{Hastings00}.

We study the spin glass model defined by the Hamiltonian
\begin{align}
\mathcal{H}=-\frac 12 \sum_{ij}J_{ij}\mathbf{s}_i.\mathbf{s}_j -
             \sum_{i}\mathbf{h}_i.\mathbf{s}_i,
\end{align}
where the $N$ spins $\mathbf{s}_i$ are vectors with $m$ components and we use
the normalization $\mathbf{s}_i^2=m$. The vectors $\mathbf{h}_i$ are
$m$-component gaussian random fields with correlator 
\begin{align}
\overline{h_i^a
h_j^b}&=h^2\delta_{ij}\delta_{ab}
\label{hcorr}
\end{align}
and field strength $h$.

In this paper our numerical work is on the mean-field limit where the
off-diagonal $J_{ij}$ are independent gaussian random variables with variance
$1/N$, and $J_{ii}=0$. Some of our results, however, also apply for $J_{ij}$
corresponding to a finite dimensional system. In all cases the partition
function at a temperature $T=1/\beta$ can be written as
\begin{align}
Z &= \int_{-\infty}^{\infty}\left(\prod_{i\alpha}ds_i^\alpha\right) 
\left(\prod_i \delta(\mathbf{s}_i^2-m)\right)
\mathrm{e}^{-\beta\mathcal{H}}\\
&= \int_{-\infty}^{\infty}\left(\prod_{i\alpha}ds_i^\alpha\right) 
\int_{-i\infty}^{i\infty}\left(\prod_i \frac{\beta dH_i}{4\pi}\right)
\nonumber\\
&\quad\times
\exp\left(\frac\beta2 \sum_i H_i (m-\mathbf{s}_i^2)+
\frac{\beta}{2} \sum_{ij}J_{ij}\mathbf{s}_i.\mathbf{s}_j
+\beta\sum_{i}\mathbf{h}_i.\mathbf{s}_i\right),
\end{align}
where we have introduced an integral representation for the $\delta$
functions.  The integrals over the spin components $s_i^\alpha$ can be done,
and making use of Eq.~\eqref{hcorr} this results in
\begin{multline}
Z = \int_{-i\infty}^{i\infty}\left(\prod_i \frac{dH_i \beta}{4\pi}\right)\\
\times\exp\left[\frac{\beta m}{2}\left(\sum_i (H_i+h^2 \chi_{ii}) +
  \frac 1\beta \ln\det(\chi/\beta) \right)\right],
\label{partition}
\end{multline}
where the matrix $\chi$ is defined by
\begin{align}
\chi_{ij} &= (A^{-1})_{ij}\qquad\text{with}\\
A_{ij} &= H_i\delta_{ij}-J_{ij}.
\end{align}
For large $m$, the integral in Eq.~\eqref{partition} can be evaluated by a
steepest descent calculation in the $H_i$, giving rise to the conditions
\begin{align}
\beta=\chi_{ii}+\beta h^2 (\chi^2)_{ii} \qquad i=1,\dots,N
\label{finiteT}
\end{align}
which determine the $H_i$.  Eq.~\eqref{finiteT} can be rewritten using the
eigenvector decomposition of $A$, the inverse of $\chi$, as
\begin{align}
\beta &= \sum_n \frac{(a^n_i)^2}{\lambda_n}
  \left(1+\frac{\beta h^2}{\lambda_n}\right),
\label{finiteTb}
\end{align}
where $\lambda_n$ are the eigenvalues and $a_i^n$ are the orthonormal
eigenvectors of $A$.

At zero temperature, Eqs.~\eqref{finiteT} and \eqref{finiteTb} are no longer
well-defined. In the ground state all spins are aligned parallel to their
local field, i.e.
\begin{align}
H_i \mathbf{s}_i &= \sum_j J_{ij} \mathbf{s}_j.
\label{zeroT}
\end{align}
For the large-$m$ limit there exists a unique stable solution of these
equations for the $H_i$, as opposed to the case of the finite $m$ spin glass
which has an exponentially large number of stable solutions
\cite{BrayMoore81}. This fact allows us to solve Eqs.~\eqref{zeroT}
numerically without difficulty. We note that it is not immediately obvious
that the $H_i$ in Eq.~\eqref{zeroT} are in any way related to the $H_i$ buried
in the matrix $\chi$ in the finite temperature problem,
Eq.~\eqref{finiteT}. It has been shown in \cite{Hastings00}, however, that the
$H_i$ in Eq.~\eqref{zeroT} are equal to the limit of the $H_i$ in
Eq.~\eqref{finiteT} as $\beta\to\infty$. This observation allows us to cover
the complete temperature range including $T=0$ within one framework.

Given the $H_i$ determined by Eq.~\eqref{finiteT} and disregarding any
irrelevant prefactors, the partition function is then
\begin{multline}
Z=\exp\left[\frac{\beta m}{2}\left(\sum_i (H_i+h^2 \chi_{ii}) +
  \frac 1\beta \ln\det(\chi/\beta) \right)\right]\\
\times\exp\left(-\frac 12\ln\det(B/\beta^2)-\frac N2\ln m\right),
\label{expansion1m}
\end{multline}
where we have included the first order fluctuation corrections around the
steepest descent values for $H_i$ which involve the matrix $B$ defined by
\begin{align}
B_{ij}&=\frac 12 \chi_{ij}^2 + \beta h^2 \chi_{ij}(\chi^2)_{ij}.
\label{Bdef}
\end{align}
This expansion is valid for any spatial dimension $d$ including $\infty$ since
we have so far made no assumptions about the matrix $J_{ij}$.

It has been shown by de Almeida et al.\ \cite{deAlmeidaEtAl78} that the
$m=\infty$ SK spin glass is replica symmetric and in the thermodynamic limit
has the same free energy as the spherical spin glass model
\cite{KosterlitzEtAl76} (but despite having the same free energy, the physics
of the low temperature phase is in fact very different from the spherical
model).  All the $H_i$ are equal in the thermodynamic limit, and the
distribution of eigenvalues of the matrix $A$ follows the Wigner semicircle
law. The phase transition (at $h=0$) then follows from the fact that
Eq.~\eqref{finiteTb} only has a solution for $0\le\beta\le \beta_c=1$, which
is
\begin{align}
H_i=\beta+1/\beta.
\label{Hsol}
\end{align}
At the critical temperature $T_c=1/\beta_c$, the smallest eigenvalue becomes
zero.  

If the field is non-zero, however, there is no phase transition since the term
involving $1/\lambda_n^2$ allows for a solution at any temperature.  This can
be demonstrated by evaluating Eq.~\eqref{finiteTb} under the assumption that
again all $H_i$ are equal to $H$ in the thermodynamic limit and thus the
Wigner semicircle law holds, which then yields after some algebra
\begin{align}
\beta^2 h^4 &= (\beta^2 + 1 + \beta^2 h^2 - \beta H)(H^2-4).
\end{align}
This equation has a positive physically relevant solution $H$ for any $\beta$,
as long as $h\ne 0$.

The scenario in finite dimensions has been described in \cite{BrayMoore82}.
This work was an extension of the Bose glass theory of Hertz, Fleishman, and
Anderson \cite{HertzEtAl79}.  For temperatures $T>T_c$ the eigenfunctions
corresponding to the smallest eigenvalues of the matrix $A$ are localized but
become extended at criticality. The $H_i$ vary from site to site. However,
Eq.~\eqref{finiteT} still applies.  The second term on the right hand side of
Eq.~\eqref{finiteT} involves the replicon susceptibility $\chi^2$. Assuming
there is a phase transition in non-zero field, this is the quantity which
diverges on the critical line \cite{BrayRoberts80}. This, however, makes it
impossible for Eq.~\eqref{finiteT} to be satisfied, thereby ruling out the
existence of a phase transition in a field, i.e.\ an Almeida-Thouless line
\cite{DeAlmeidaThouless78}, by contradiction.  As this line marks the onset of
replica symmetry breaking, we deduce from its absence that the large $m$ limit
and the straightforward expansion in $1/m$ about it, produces a theory which
is replica symmetric.  The absence of such a line is a prediction of the
droplet theory of spin glasses. However, usually with $1/m$ expansion methods
one does not adopt the direct expansion approach in powers of $1/m$ but
instead uses (ad-hoc) self-consistent approximations, such as could be
generated by making the sum of the terms in $m$ and of $O(1)$ in the
exponentials in Eq.~\eqref{expansion1m} stationary with respect to the $H_i$,
and it is possible that with such an approach replica symmetry breaking might
emerge.

We turn now to a more detailed study of the phase transition mechanism at zero
field in the SK limit.  First we examine the ground state properties in zero
field. Eq.~\eqref{zeroT} can be written as
\begin{align}
\sum_j\left(H_i\delta_{ij}-J_{ij}\right)s_j^\alpha &=0,
\label{zeroTb}
\end{align}
where $\alpha=1,\dots,m$ labels the spin components, showing that the matrix
$A_{ij}=H_i\delta_{ij}-J_{ij}$ has \textit{at least} one eigenvalue which is
\textit{exactly} 0. On the other hand, it was shown in \cite{Hastings00} that
there is an upper limit of the number of null eigenvalues $n_0$, which is
$n_0<\sqrt{2N}$.

The number of null eigenvalues will turn out to be the key quantity of the low
temperature phase of the large-$m$ spin glass. Here, we can already see some
implications. The fact that there are $n_0$ null eigenvalues reduces the
\textit{effective} number of spin components from $m$ to $n_0$. To see this,
consider the matrix formed by the entries of the spins, $s_i^\alpha$. Since
the rows of this matrix (regarding $\alpha$ as the row index) correspond to
null eigenvectors of $A$ and there are only $n_0$ linearly independent ones,
the row rank of this matrix is at most $n_0$. But since the row rank and the
column rank of any matrix are equal, the column rank is also at most
$n_0$. Therefore the $N$ columns, being the spins, can only span a
$n_0$-dimensional subspace of the $m$ dimensional space they live in. Making
use of the global rotation invariance of the spins, it is therefore sufficient
to set $m=n_0$, or at least $m=\sqrt{2N}>n_0$ when $n_0$ is yet
unknown. Obviously, this accelerates numerical simulations, allowing one to go
to relatively large system sizes, as we will see in the following.

\begin{figure}[htb]
\centerline{\epsfig{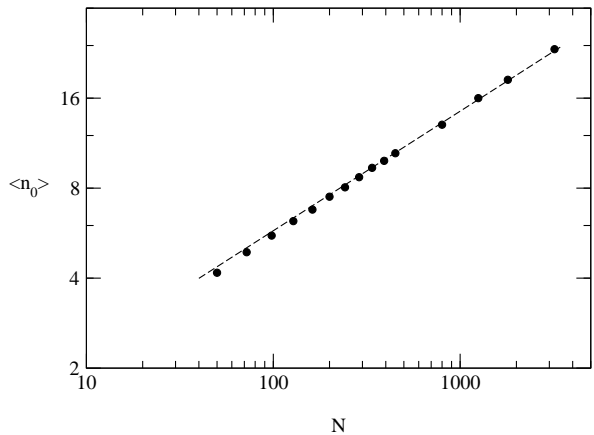}}
\caption{The average number of null eigenvalues $\langle n_0\rangle$ as a
function of system size $N$. The error bars on the data points are smaller
than the point size. The dashed line is $\sim N^{2/5}$ for comparison.}
\label{zeroTfig}
\end{figure}

\begin{figure}
\epsfig{file=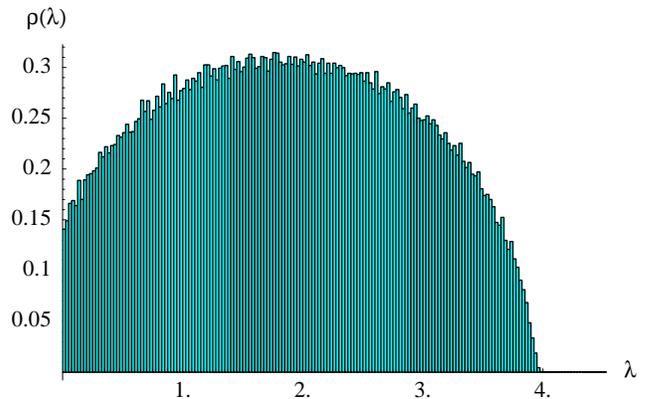,width=\columnwidth}
\caption{Density of eigenvalues of the matrix $A$ for a system with $N=450$
  (but typical for all system sizes), excluding the null eigenvalues, averaged
  over 150 samples. The density $\rho(\lambda)$ goes to a constant at small
  $\lambda$. (The constant tends to zero for large system sizes, and for an
  infinite system the Wigner semicircle is restored.) }
\label{densityofstates}
\end{figure}

We have solved Eqs.~\eqref{zeroT} numerically for the spins $\mathbf{s}_i$ by
straightforward iteration, setting $H_i=|\sum_j J_{ij}\mathbf{s}_j|/\sqrt{m}$
at each step, until the average angular deviation of the spins from their
local field directions was smaller than some prescribed accuracy (we used
$10^{-8}$). From the ground state configuration found in this way we
determined the matrix $A$ and analysed its eigenvalues. We found that the
smallest non-null eigenvalues are always at least 5 orders of magnitude larger
than the null eigenvalues (which we find numerically to be of the order
$10^{-8}$ or smaller). The average number of null eigenvalues as a function of
the number of spins is plotted in Fig.~\ref{zeroTfig}.
In the range of system sizes accessible to us, it behaves as $n_0\sim
N^{2/5}$.  This agrees with the prediction from \cite{Hastings00}. The
calculation leading to this result was, however, based on an approximation of
the density of states of the matrix $A$ which, at finite values of $N$, had a
gap in the density of states and a square root singularity at the edge of the
gap. Numerically we found a very different behavior, see
Fig.~\ref{densityofstates}. Instead of a gap there is an \textit{enhancement}
in the density of states at low eigenvalues (rather similar to the behavior
produced in the three-dimensional XY spin-glass model by finite size effects
\cite{BrayMoore82b}).

A new argument is therefore needed to explain the observed value of the
exponent $2/5$. From \cite{deAlmeidaEtAl78,BrayMoore81} it can be deduced that
the ground state energy per spin of the $m$-component spin glass, divided by
$m$, goes as $-1+1/4m+\mathcal{O}(1/m^2)$. Since $n_0$ is equal to the
effective number of spin components, the system would at first sight be able
to attain its lowest energy state by choosing $n_0$ as large as possible,
i.e.\ equal to its upper bound. However, this calculation of the energy was
done taking the thermodynamic limit first, such that $m$ (or $n_0$) is always
much less than any power of $N$. When $n_0$ is comparable to some power of
$N$, there are additional energy costs, whose magnitude can be estimated by
the following argument. Starting from all $H_i$ being equal and then tuning
them in such a way that there are $n_0$ null eigenvalues will result in a
downward shift in the eigenvalue spectrum of $A$ which is of order
$(n_0/N)^{2/3}$ (the lowest $n_0$ eigenvalues in a Wigner semicircle reach
this far from the band edge, and the shift is expected to be of the same
order). The ground state is reached when the energy arising from these two
competing terms $1/4n_0+\mathrm{const.}(n_0/N)^{2/3}$ is minimized, which is
the case when $n_0\sim N^{2/5}$.

It is striking that in the ground state the individual spins condense into a
$n_0$-dimensional subspace of their original $m$-dimensional space. This
behaviour is a generalization of the conventional Bose-Einstein condensation
where the constituents condense into a \textit{single} (one-dimensional)
state, which is  also what happens in the spherical spin glass model 
\cite{KosterlitzEtAl76}.

To show that the behavior we observed at zero temperature is in fact the
generic behaviour in the low temperature phase, we solved Eqs.~\eqref{finiteT}
numerically in zero field. Since there is no phase transition for a finite
number of spins, we were able to solve Eqs.~\eqref{finiteT} numerically over a
very large temperature range, both in the high and low temperature regions.
We used a standard Newton-Raphson iteration scheme for this purpose. This
method converges very quickly for high temperatures $\beta<1$ and fails to
converge for low temperatures unless the starting configuration for the
iteration is already sufficiently close to the solution. In order to ensure
this, we employed the exact differential equation
\begin{align}
\frac{dH_i}{d\beta} &= -\frac 12 \sum_j (B^{-1})_{ij},
\end{align}
with $B$ as defined in Eq.~\eqref{Bdef} (with $h=0$).  This equation, which
can easily be derived from Eq.~\eqref{finiteT}, was used to project from a
solution found at $\beta$ to a good initial configuration at
$\beta+\Delta\beta$. Using this method we were able to track the solution of
Eq.~\eqref{finiteT} over the temperature range $\beta=0.1\dots 100$.

\begin{figure}
\centerline{\epsfig{file=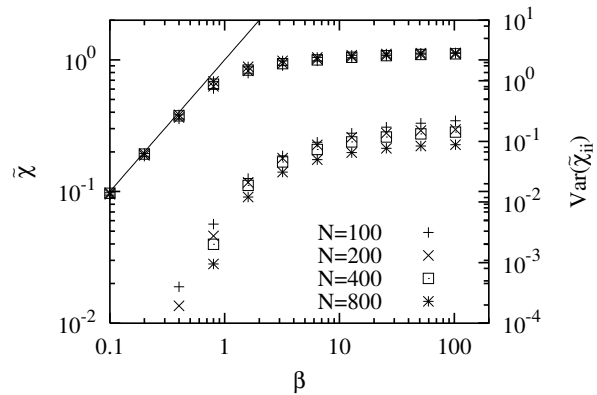,width=\columnwidth}}
\caption{The site-averaged susceptibility $\tilde{\chi}$ (upper curves,
  left axis) and its variance (lower curves, right axis). The solid line
  corresponds to $\tilde{\chi}=\beta$ for comparison. }
\label{splitfig}
\end{figure}

We have found it useful to split Eq.~\eqref{finiteTb} at $h=0$ into
two parts corresponding to the eigenvalues that are going to zero as
$\beta\to\infty$ and the rest,
\begin{align}
\beta &= \sum_{\lambda_n\text{ going to 0}} \frac{(a_i^n)^2}{\lambda_n} +
  \sum_{\lambda_n\text{ staying finite}} \frac{(a_i^n)^2}{\lambda_n}.
\label{spliteq}
\end{align}
In a finite system at finite temperature, all eigenvalues are naturally
non-zero, such that this equation is well-defined. However, for
$\beta>\beta_c$ the eigenvalues which become exactly 0 at $T=0$ decrease with
system size as $N^{-b}$ at finite temperature, where $b$ is an unknown
positive exponent, and so are equal to 0 in the thermodynamic limit throughout
the whole low temperature phase. In this situation, the second sum in
Eq.~\eqref{spliteq} is equal to the diagonal elements of the Moore-Penrose
inverse (see, e.g., \cite{Ben-IsraelGreville74}) of $A$ which corresponds to
the physical susceptibility $\tilde{\chi}_{ij}= \partial \langle
s_i^\alpha\rangle/\partial h_j^\alpha = \beta(\langle s_i^\alpha
s_j^\alpha\rangle - \langle s_i^\alpha\rangle\langle s_j^\alpha\rangle)$. For
$\beta>\beta_c$ Eq.~\eqref{spliteq} can therefore be written as
\begin{align}
\beta=\beta\langle s_i^\alpha\rangle^2 +
\tilde{\chi}_{ii},
\label{eachi}
\end{align}
while for $\beta<\beta_c$, $\tilde{\chi}_{ii}=\chi_{ii}=\beta$.  In
Fig.~\ref{splitfig} we have plotted $\tilde{\chi}$, defined as
$\tilde{\chi}_{ii}$ averaged over sites, and its variance
$\mathrm{Var}(\tilde{\chi}_{ii})$ as a function of $\beta$. The plot shows
that above the transition temperature $\tilde{\chi}$ is equal to $\beta$,
whereas below the transition temperature $\tilde{\chi}$ remains essentially
frozen at a value of $\beta_c$.  The site-to-site variance of $\tilde{\chi}$
decreases with system size, roughly following a power law
$\mathrm{Var}(\tilde{\chi}_{ii})\sim N^{-1/3}$ (data not shown here).
Eq.~\eqref{eachi} then implies that the Edwards-Anderson order parameter
$q=\langle s_i^\alpha\rangle^2$ is equal to $1-\beta_c/\beta=1-T/T_c$, i.e.\
the frozen components of the spin are those associated with the null
eigenvalues of the matrix $A$.

We believe that the approach to spin glasses via a $1/m$ expansion method,
while it is not simple to carry out, seems to be the only calculational
technique which can avoid the problems associated with the unwanted
\cite{NewmanStein03} replica symmetry breaking of the loop expansion. The
point about which the expansion takes place, viz the large-$m$ limit, is also
interesting in its own right as the unusual phase transition mechanism found
in the SK limit would be expected to carry over to finite dimensions. The work
by Viana \cite{Viana88}, who found that in the large-$m$ limit the upper
critical dimension and the lower critical dimension are both equal to 8,
indicates that below the upper critical dimension there is no Edwards-Anderson
order but instead perhaps chiral order as suggested by Kawamura
\cite{HukushimaKawamura00,ImagawaKawamura03} (but contested by
\cite{LeeYoung03}). Our approach might in the future be developed into a tool
to study this controversy from a new perspective.

\begin{acknowledgments}
We thank M.\ B.\ Hastings for valuable discussions.  TA acknowledges financial
support by the German Academic Exchange Service (DAAD) and the European
Community's Human Potential Programme under contract HPRN-CT-2002-00307,
DYGLAGEMEM.
\end{acknowledgments}

\bibliography{Spinglass} 

\end{document}